# Superconducting bimodal ionic photo-memristor

Ralph El Hage[1]. Vincent Humbert[1], Victor Rouco[1], Anke Sander[1], Jérôme Charliac[2], Salvatore Mesoraca[1], Juan Trastoy[1], Javier Briatico[1], J. Santamaría[1,3] & Javier E. Villegas[1,*]

[1]*Unité Mixte de Physique, CNRS, Thales, Université Paris Saclay, 91767 Palaiseau, France*

[2] *Laboratoire de Physique des Interfaces et des Couches Minces (UMR7647), CNRS, Ecole Polytechnique, 91128 Palaiseau Cedex, France*

[3]*GFMC, Dpto. Física de Materiales. Universidad de Ciencias Físicas, Universidad Complutense de Madrid, 28040 Madrid, Spain*

Memristive circuit elements constitute a cornerstone for novel electronic applications, such as neuromorphic computing, called to revolutionize information technologies. By definition, memristors are sensitive to the history of electrical stimuli, to which they respond by varying their electrical resistance across a continuum of nonvolatile states. Recently, much effort has been devoted to developing devices that present an analogous response to optical excitation. Here we realize a new class of device, a tunnelling photo-memristor, whose behaviour is bimodal: both electrical and optical stimuli can trigger the switching across resistance states in a way determined by the dual optical-electrical history. This unique behaviour is obtained in a device of ultimate simplicity: an interface between a high-temperature superconductor and a transparent semiconductor. The microscopic mechanism at play is a reversible nanoscale redox reaction between both materials, whose oxygen content determines the electron tunnelling rate across their interface. Oxygen exchange is controlled here via illumination by exploiting a competition between electrochemistry, photovoltaic effects and photo-assisted ion migration. In addition to their fundamental interest, the unveiled electro-optic memory effects have considerable technological potential. Especially in combination with high-temperature superconductivity which, beyond facilitating the high connectivity required in neuromorphic circuits, brings photo-memristive effects to the realm of superconducting electronics.

* javier.villegas@cnrs-thales.fr

The search for faster, energy-efficient memories and novel computation schemes has fostered the exploration of resistive switching effects[1]. Observed in a variety of systems that span from magnetic[2,3] or ferroelectric[4,5] tunnel junctions to transition-oxide capacitors[6,7] and strongly correlated materials[8–10], the term resistive switching denotes a "jump" between non-volatile electrical resistance states (high/low resistance, 0/1 for logics) generally triggered by a voltage or a current pulse.

Memristors[11] constitute a particular class of two-terminal resistive switching devices whose functionality is beyond that of a binary memory: they show a continuum of states - instead of only two - and the switching between them is driven by the time-integrated current across the device (or by the history of applied voltages). Thus, one can think of memristors as multi-state memories switchable by cumulative stimuli. Depending on the resistance states' lifetime and dynamics, memristors can mimic the function of either synapses or neurons[11,12], thus constituting a cornerstone for the nascent field of neuromorphic computing[13–16].

A related, tantalizing idea is the development of memristors sensitive to light[17–25] —particularly, in which illumination can trigger a switching across non-volatile resistance states. This type of optical memory may be game-changing in applications such as photonic neural networks[26] and open the door to novel optoelectronic devices[27,28] —for example, neuromorphic vision sensors[21,29]. Various systems have been recently explored in search of photomemristive effects. Those include complex nanostructures based on optically active polymers[17], metal-oxide capacitors[21,22,24], all-oxide[19] and semiconductor[18] heterostructures, as well as ferroelectric tunnel junctions[23]. Their conductance shows photosensitivity due to mechanisms that span from to light-induced polymer contraction/expansion[17] and electron trapping/detrapping[18,19,21,22,24] to photoinduced ferroelectric switching[23]. In those systems, the resistance states' lifetimes range from typically seconds in electronic processes[18,19,21,22,24] to minutes in organic systems[17], up to the virtual non-volatility of ferroelectric devices[23].

In addition to the common challenges associated with memristors, e.g. obtaining large resistance variations to facilitate readout and finding geometries enabling the high connectivity required for neuromorphic circuits[11,14–16], photo-memristors pose additional specific ones. Namely, in memristors electrical excitation often allows for bidirectional switching —the polarity of the electrical stimulus determines whether the resistance increases or decreases. Such property is crucial for mimicking the so-called depression and potentiation phenomena characteristic of synaptic plasticity[14,15,17,18]. An analogous function is generally absent in the optical response (in most cases illumination only produces a resistance decrease[19,21–23]) except for a few realizations that require the combination of various light sources. For instance, in polymer-based devices the opposite effects of circularly and linearly polarized light respectively lead to a resistance increase or decrease, thus mimicking depression and potentiation[17]. Illumination under variable wavelength has also been exploited to that end, particularly in systems based on electron trapping/detrapping[24], in which the natural relaxation of visible-light excited conductance states is potentiated via infrared illumination, thus allowing for a virtually bidirectional optical switching.

Here we report on a new class of photo-memristor that exploits a distinct microscopic mechanism: a controllable oxygen exchange between the two materials that constitute the device —a junction between a superconducting cuprate and a semiconducting oxide. Crucially, such a mechanism allows for giant resistance switching effects, yields virtually non-volatile states, and can be driven both optically and electrically. Indeed, a remarkable specificity of our ionic photo-memristor is that its response to a given optical stimuli depends on the electrical history. Due to this entanglement, and at variance with other approaches, a single light source can controllably produce bidirectional switching. This behaviour results from an unusual, competing interplay between electrochemistry, photon-activated oxygen diffusion and strong photovoltaic effects. Interestingly, the present demonstration is based on a high-temperature superconducting cuprate. As further discussed below, this greatly multiplies its technological potential, not only because superconductivity enhances resistive switching effects and facilitates the high-connectivity required for neuromorphic circuit, but also because electro-optical memory is a novel, game-changing function in the thriving field of superconducting electronics[30–35].

The key specificities of the electro-optical switching behaviour observed here are schematically summarized in Figure 1. The application of voltage pulses $V_{write}$ (of the order of a few Volts) enables a conductance switching $\Delta G_E$ across a continuum of non-volatile levels, whose non-destructive readout is possible with a much lower $V_{read}$ of the order of mV. The conductance levels span between two extreme states —thereafter called ON (high conductance) and OFF (low conductance) — that can be up to four orders of magnitude afar. The electrical switching is hysteretic, bidirectional, and reversible. This behaviour is qualitatively and quantitatively similar to the acclaimed tunnel electroresistance (TER) of ferroelectric tunnel junctions[23,36–38]. Strikingly, illumination with visible or UV light also leads to a conductance switching across non-volatile states, and it does so in a very characteristic fashion: the conductance level shifts in opposite directions depending on the previous electrical junction's state. Namely, optical stimuli lead to either an enhancement or a decrease of the conductance $\Delta G_{Op}$ depending on whether the device had been previously set in (or nearby) the OFF or ON state. While this the electric history determines the sign of $\Delta G_{Op}$, crucially the amplitude of $\Delta G_{Op}$ cumulatively depends on the number of photons shone on the device, that is, it is controlled by the optical history. In addition to the electro-optical switching summarized in Fig. 1, the devices studied here show unusually large photovoltaic effects which, as discussed thereafter, play a key role in the photo-memristive behaviour.

The scheme of the superconducting tunnelling photo-memristor is shown in Fig. 2a. It consists of a micrometric junction between the archetypal high-$T_C$ superconductor $YBa_2Cu_3O_7$ (YBCO, bottom electrode) and the transparent semiconductor indium-tin-oxide (ITO, top electrode). The junctions are fabricated on 30 nm thick c-axis oriented YBCO films grown epitaxially on (001) $SrTiO_3$ (STO) substrates. A ~ 1 µm-thick photoresist is spin-coated on the YBCO film and micrometric openings are photolithographed. The amorphous ITO subsequently deposited on the resist contacts the YBCO surface across those openings, forming the junctions (for further details see[39] and Methods). This simple layout allows measuring differential conductance $G \equiv dI/dV_{read}$ across the YBCO/ITO interface using the electrical wiring sketched in Fig. 2a.

The electrically-driven switching of the tunnel conductance $\Delta G_E$ originates from a reversible redox reaction between ITO and YBCO that greatly changes the interface's transparency to electrons and hence the electron tunnelling rate across it. This is as demonstrated earlier in YBCO/Mo$_{80}$Si$_{20}$ junctions and sketched in the cartoon of Fig. 2a[39]. The OFF state corresponds to the junctions' virgin and ground state. As dictated by the high reduction potential of Cu in YBCO, compared to that of Sn or In in ITO (Supplementary Table 1), a spontaneous reduction of the interfacial YBCO occurs upon deposition of ITO. Thus, within the first 2-3 nm from the interface, YBCO becomes severely oxygen-depleted and insulating[39]. As one moves away from the interface the oxygen content increases gradually, which results in a first YBCO region with depressed superconducting properties (low T$_c$ YBCO) then to optimal oxygenation and T$_C$ further away (scheme in Fig. 2a).[39] The insulating YBCO layer behaves as a tunnelling barrier, and its presence leads to the OFF state –hollow-symbols curve in Fig. 2b. The application of a sufficiently high $V_{write} > 0$ can reverse the redox reaction, so that oxygen is driven back into the interfacial YBCO, thinning down both the insulating and low-T$_C$ YBCO layers.[39] This increases the junction conductance by orders of magnitude, leading to the ON state –solid symbols curve in Fig 2b. The junction can be reversibly switched from ON to OFF and *vice versa*, crossing over a continuum of intermediate states, by applying negative/positive $V_{write}$ (in the few volts range). This is shown in Figure 2c, which displays the remnant conductance for $V_{read} = 100$ mV and $V_{read} = 0$ mV (labelled $G_{100}$ and $G_0$) following the application of different $V_{write}$ at $T = 4K$ across a loop (the chirality is indicated by the spinning arrow). Similarly as in earlier experiments[39], the hysteretic switching is fully reproducible (two $G_{100}$ and two $G_0$ loops are superposed in Fig. 2c) and asymmetric. In particular, switching from ON to OFF requires a smaller $|V_{write}|$ than the opposite ($V_{write}$~ $-1$ V *vs.* $V_{write}$~2.5 V). The highest electroresistance $ER \equiv G_{ON}/G_{OFF}$ is observed for low $V_{read}$: $ER_{0\ mV}$~540 while $ER_{100\ mV}$~100. The temperature dependence of the conductance, shown in the inset of Fig 2b for the ON and OFF states, also behaves differently at low and high $V_{read}$. At $V_{read} = 100$ mV (circles), a monotonous trend is observed across the entire temperature range: $G_{ON}$ is nearly constant and $G_{OFF}$ logarithmically decreases with decreasing temperature. However, for $V_{read} = 0$ mV, a departure from those trends is observed below the superconducting critical temperature T$_C$ (which is lower in the OFF than in the ON state). As we demonstrated earlier[39], the greater $ER$ and distinct temperature behaviour observed for $V_{read} = 0\ mV$ are explained by the opening of the superconducting gap, which reduces the electronic density of states at the Fermi level and dramatically decreases the tunnelling conductance. In summary, the YBCO/ITO junction presents similar tunnelling electroresistance effects as those found earlier in YBCO/MoSi junctions, including a pronounced electroresistance enhancement in the superconducting state.

While the OFF state is the junctions' ground state, and therefore is stable over time, the ON state is metastable and, at sufficiently high temperatures, it relaxes over time $t$ towards the OFF state. This is demonstrated in Fig. 2d, which shows the time-dependent $\Delta G_E(t)/G_{ON}(t=0)$ at different temperatures. $\Delta G_E$ stays constant as a function of time at low temperatures (up to ~ 90 K - 100 K), but relaxes over time at higher ones, at a rate that increases with increasing temperature.

We detail in what follows the optical effects, which are divided in two classes. The first one corresponds to a non-volatile effect: a photoinduced increase or decrease of the junction's conductance that is persistent once in the dark (as long as the temperature is kept below ~90 K - 100 K). The second one corresponds to a volatile effect, and consists of a photovoltage that disappears as soon as illumination is halted.

The first type of effects, i.e. the optical switching between non-volatile states, is summarized in Fig. 3. Figs. 3a and 3b respectively display the differential conductance $G(V_{bias})$ for a same junction after it has been set in the ON and OFF state and subsequently illuminated with visible light ($\lambda = 405 nm$) over a time ranging from 15 to 120 minutes (see labels). Strikingly, the photo-response is very different in each case. If the junction is set in the ON state prior to illumination (Fig 3a), a gradual photo-induced decrease of conductance is observed as the illumination time increases (series of blue shaded curves). Notice that for the longest illumination time (120'), the overall conductance drops by a factor of ten. Contrarily, if the junction is set in the OFF state prior to illumination (Fig. 3 b), a gradual increase of the conductance is observed as the illumination time increases. It is important to stress that the conductance levels set by illumination are non-volatile as long as the junction is maintained in the dark and at temperatures below ~ 90 K - 100 K. Furthermore, and whatever the conductance level reached after illumination, the junctions can be reset to the ON/OFF state by the application of a positive/negative $V_{write}$ (as shown for example by the green curves in Fig. 3a and 3b).

Fig. 3c displays the relative change of the conductance in the ON and OFF state, $G_{light}/G_{ON}$ and $G_{light}/G_{OFF}$, as a function of the illumination time and for different optical powers (indicated by the labels), with $G_{light}$ (and $G_{ON}$ or $G_{OFF}$) the conductance after (prior to) illumination. Data are extracted for $V_{read} = 100$ mV from measurements as those in Fig. 3a and 3b. One can see that light effects are proportional to the illumination time and power. That is, the devices behave as a photon integrator: the size the photoinduced conductance change depends on the number of photons shone on the device.

In summary, whether illumination produces an increase or a decrease of the conductance level depends on the previous electrical history, and the size of the conductance variation depends on the number of photons shone on the device (optical history). The optical and electrical memristive responses are thus coupled, which results in a new functionality in which light effects are bidirectional, cumulative and can be "erased" by application of a voltage pulse.

The photoinduced effects in the ON and OFF states also differ in their temperature dependence. This is shown in Fig. 3d, which displays the relative change of the conductance after illumination ($G_{light}/G_{ON}$ and $G_{light}/G_{OFF}$ at $V_{read} = 100$ mV) as a function of temperature, for various samples and illumination conditions (see figure caption). One can see that, in all cases, in the OFF state (red data points) the photo-induced changes are the largest (up to 1000%) at low temperature and gradually decrease as temperature is increased, until they become only a few percent at around 100 K. Contrarily, in the ON the state the effects are relatively small at low temperature and gradually increase as temperature is increased, reaching up to 5000% at ~ 100 K. Notice that there exists an intermediate range of intermediate temperature where both effects are similarly strong. The completely different temperature dependence suggests that optical switching arises in the ON and OFF states from different microscopic mechanisms.

Fig. 4 illustrates the second class of photoinduced phenomena, namely the photovoltaic effects. Fig, 4a shows the typical current *I* vs $V_{bias}$ junction measured in the OFF state, both in the dark (black curve) and under illumination with $\lambda = 405$ nm and $P \approx 150 \ mW.cm^{-2}$ (blue curve). As soon as the junction is illuminated, we observe a large photovoltaic effect, i.e. a shift downwards of the *I*-$V_{bias}$ curve. This is caused by the appearance of a photocurrent that instantly disappears when the illumination is halted (see arrow). We define the short-circuit current $J_{sc}$ as the current measured when a zero bias is applied across the junction, and the open circuit voltage $V_{oc}$ as voltage across the device when no electrical current is measured. While $V_{oc}$ (~0.5V) is comparable to that observed earlier in cuprate based photovoltaic cells[40,41], $J_{sc}$ is several orders of magnitude higher[40,41] and indeed compares to that of silicon based devices[42]. Figure 4b displays $V_{oc}$ as a function of the junction's remnant conductance measured at $V_{bias} = 100 \ mV$, $G_{100}$. This conductance level is set prior to illumination by application of $V_{write}$ pulses across the hysteretic switching loop (inset of figure 4b). The magnitude of $V_{oc}$ strongly depends on $G_{100}$, that is, on whether the sample is in the ON, OFF or intermediate states. In particular, one can see that the photovoltage is highest at low conductance (OFF state), and logarithmically decays as the junction conductance is increased towards the ON state. Notice that $V_{oc}$ depends only on the conductance state $G_{100}$ and not on the electrical history, since the data points from the ON/OFF and OFF/ON switching branches (respectively solid and hollow symbols) collapse into a single master curve.

In order to describe the model that accounts for all the experimental observations, let us start by discussing the origin of the photovoltaic effect demonstrated in Fig. 4. For this we need to recall the interface structure displayed in Fig. 2a, which explains the resistance switching effects based on the presence of a strongly oxygen-depleted YBCO interfacial region that plays the role of tunnelling barrier. Strongly oxygen-depleted YBCO is in fact a p-type semiconductor with a ~1.25 eV gap[43] and the adjacent ITO is a degenerate n-type semiconductor with a ~3.5 eV gap [44]. In this sense, the interface can be seen as a p-n junction and a photovoltaic effect is therefore expected. The different work function in both materials, $\phi_{ITO} = 4.5$ eV[45] and $\phi_{YBCO} \approx 5.5 - 6$ eV[46], leads to a space charge layer (SCL) with a built-in electric field pointing towards the YBCO, as it is sketched in figure 5a. In this scenario, upon illumination, electrons and holes photogenerated in the absorber YBCO respectively flow towards the YBCO and ITO, leading to the observed photocurrent (photovoltage in open-circuit configuration). As a consistency check, we can obtain a rough estimate of the expected SCL width using[47] $W = \sqrt{(N_A+N_D/N_AN_D)\,2\varepsilon_s V_{bi}/e}$ where $N_A \sim 8\,10^{19} cm^{-3}$ and $N_D \sim 5\,10^{20} cm^{-3}$ are respectively the carrier densities in oxygen-depleted YBCO[48] and in ITO[49], and an upper limit for built-in voltage $V_{bi} < 1.5V$ can be inferred from the difference between work functions. From this, $W < 5nm$ is obtained which is comparable with the range of thickness of the severely oxygen-depleted YBCO layer at the interface[39]. The scaling of $V_{oc}$ with the logarithm of the junction's conductance state $G_{100}$ observed in Fig. 4b is also consistent with the discussed scenario. Indeed, if one considers that the current across the junction and the conductance are roughly proportional in the low $V_{bias}$ regime, the behaviour of Fig. 4d is as expected from the Shockley relation[50] $V_{oc} = \frac{k_bT}{\eta q}\ln(\frac{I_{sc}}{I_0} - 1)$ where $\eta$ is the ideality factor, $q$ is the electric charge, $I_{sc}$ is the photocurrent and $I_0$ is the reverse saturation current in dark. This explains why $V_{oc}$ is the largest in the low conductance (OFF) state, in which $I_0$ is the lowest, as well as the logarithmic decay of $V_{oc}$ as the conductance (and $I_0$) increases.

Let us now discuss the optical switching effects. When the junction is set in the ON state, after illumination we observe a strong, virtually non-volatile conductance decrease. This can be explained by recalling that the system naturally tends to relax from the ON (metastable state) into the OFF (ground state), as demonstrated by the data in figure 2d. While at low temperature (below ~ 100 K) and in the dark this relaxation is very slow (in those conditions the ON state is virtually non-volatile), illumination activates the relaxation, leading to a strong conductance decrease that is stopped as soon as illumination is halted. At the microscopic level, this happens in the way sketched in Fig. 5c. In the ON state, the junction consists of an optimally doped YBCO layer separated from oxygen depleted ITO by very thin (nanometric)[39] oxygen-deficient (insulating) YBCO layer. Therefore, the system is not in equilibrium from an electrochemical point of view since the material with lowest reduction potential (ITO) presents a strong oxygen deficiency near the junction interface while most of top layers of YBCO (who has the highest reduction potential) are fully oxygenated. In this situation, light activates the diffusion of oxygen ions over the barrier, thereby acting as a catalyst for the redox reaction that naturally drives the system into its ground state. As oxygen migrates "up" from YBCO into ITO due to optical activation, the insulating oxygen-deficient YBCO interfacial layer thickens, leading to the decrease of the electron tunnelling conductance observed in Fig. 3a. It is important to stress that we ruled out that the enhanced oxygen ion diffusion observed upon illumination may be merely explained by an increase of the sample temperature. This indicates that photons directly activate ion diffusion, as proposed earlier in the literature of cuprates[51,52].

When the junction is set in the OFF state, a strong, virtually non-volatile increase of the conductance is observed after illumination. This is surprising and seemingly contradictory to the above arguments –namely that light activates the relaxation of the system into its ground state. Contrarily, here illumination drives the system towards the ON state and thus away from its ground state. This paradox can be understood of one considers two key ingredients characteristic of the OFF state. First, and contrary to the ON state, the ITO layer is optimally oxygenated and does not accept more oxygen (see sketch in Fig. 5b). Second, upon illumination, a large photovoltage is observed in the OFF state, while this is absent in the ON state (see Fig. 4b). In fact, this photovoltage drives the unexpected optical switching observed in the OFF state. As discussed above and sketched in Fig 5a, the photovoltage reflects a hole accumulation at the YBCO side of the space charge layer (SCL). This positive charge accumulation promotes the re-oxidation of the YBCO layer at the interface by attracting $O^{2-}$ ions, analogously as in well-known photocatalytic effects in which photocarriers accelerate a redox reaction in the vicinity of pn-junctions[53,54]. This local re-oxidation is made at the expense of pumping $O^{2-}$ ions from the in-depth YBCO layer that presents higher oxygen content, as sketched in Fig. 5b. In other words, the positive charge accumulation resulting from the photovoltaic effect attracts oxygen ions towards the interfacial, oxygen-depleted YBCO. Reoxygenation leads to a thinning of this layer, which is the junction's electron tunnelling barrier, thus enhancing its electrical conductance (Fig 3b). Let us stress that, crucially, because ITO is already highly oxygenated in the OFF state, oxygen diffusing from in-depth the YBCO layer does not get into the ITO, but only re-oxidizes the interfacial YBCO.

To summarize, light makes oxygen ions migrate "upwards" from the YBCO layer and across the device, both in the ON and OFF states. This migration is favoured by photon-activation over the barrier for ion diffusion and driven either by i) the redox reaction between ITO/YBCO when the junction is closer to ON state ii) by the photovoltage and ensuing hole charge accumulation, which is stronger near the OFF state. The predominance of i) or ii) depends on temperature and, crucially, on the interface's state prior to illumination –particularly on the distribution of oxygen atoms and vacancies within YBCO and ITO. This dictates whether up-streaming oxygen ions cross the interface from YBCO into ITO (leading to thickening of the tunnelling barrier and thus to a conductance decrease) or accumulate in the interfacial YBCO (leading to a thinning of the tunnelling barrier and thus to a conductance increase).

Aside from their fundamental interest, the described effects are also technologically relevant for various reasons. In addition to a unique photo-memristive behaviour, the devices studied here –a single interface– are of ultimate simplicity when compared to approaches that involve more complex geometries[17,20] and photo-active materials[17,23]. Furthermore, high-temperature superconductivity amplifies their technological potential. While the switching effects are observed at any temperature, they become much stronger below the superconducting critical temperature. This, together with the vanishing Joule dissipation, should facilitate miniaturization and the layout of dense memristor arrays with high interconnectivity, as required for neuromorphic computing[13,15]. Beyond this area, the photo-memristors realized here could be naturally implemented in superconducting electronics[30], a thriving field particularly after the advent of Josephson devices based on high-temperature cuprates[31–33] as the one used here, which allow operation above liquid nitrogen temperatures.

One could for instance cite applications such quantum antennas[34] and logic circuits[35], in which non-volatile memory is a grail[55] and photo-sensitivity was up to now unavailable. The incorporation of these functions should refashion and greatly broaden the range of applications of such Josephson circuits.

**Ackowledgements**

Work supported by ERC grant N° 647100 "SUSPINTRONICS", ERC grant Nº 966735 "SUPERMEM", French ANR grant ANR-17-CE30-0018-04 "OPTOFLUXONICS", COST action "Nanocohybri" and Spanish AEI PID2020-118078RB-I00. JS thanks a the D'Alembert program funded by the IDEX Paris-Saclay, ANR-11-IDEX-0003-02, for financing a stay at Unité Mixte CNRS/Thales. We (JS, J-E V) acknowledge funding from Flag ERA ERA-NET To2Dox project.

**Authors' contributions**

The experiments were conceived by J.E. Villegas with inputs from R. El Hage, A. Sander and J. Santamaria. The materials were grown by A. Sander, J. Briatico, and J. Charliac. The junctions were lithographed by R. El Hage, with support of S. Mesoraca. The transport measurements and data analysis were carried out by R. El Hage, V. Humbert and V. Rouco. All the named authors and J. Trastoy discussed the experimental results. The microscopic model of the photo-response was put together by R. El Hage, J. Santamaría and J. E. Villegas. The paper was written by R. El Hage and J.E. Villegas with inputs from all other the authors. The overall project was supervised by J.E. Villegas.

**FIGURES**

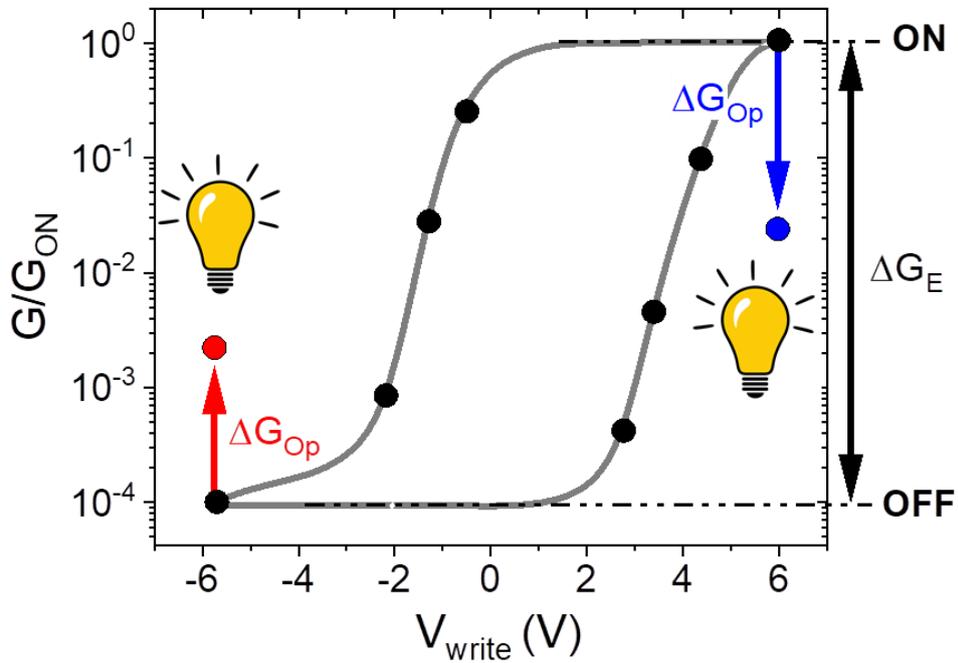

Figure 1: **Dual optical-electrical conductance switching.** Sketch of a typical hysteresis loop showing the differential conductance measured on YBCO/ITO tunnel junctions as a function of the writing pulse voltage. The black arrow indicates the maximum amplitude of the electrical switching $\Delta G_E$. The red and blue arrows respectively indicate the amplitude the optical switching $\Delta G_{Op}$ in the OFF and ON states respectively. Optical stimuli drive the junction towards the state opposite to that set electrically (ON → OFF or *vice versa*).

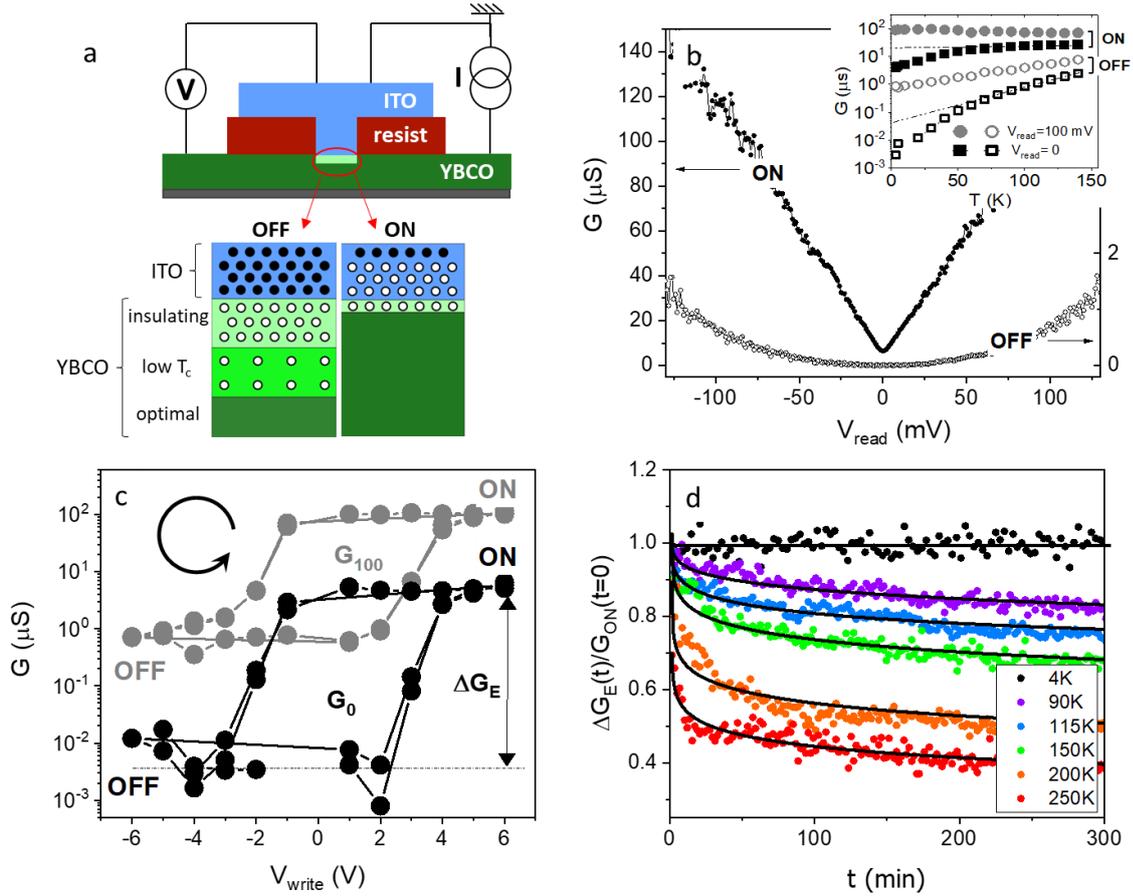

Figure 2: **Junction's scheme and electrical switching. a** Scheme of the photo-memristor composed of a c-axis $YBa_2Cu_3O_{7-\delta}$/ITO tunnel junction grown on an $SrTiO_3$(001) substrate. The micrometric junction is defined by an opening in the insulating photoresist across which the ITO is deposited by sputtering. Sketches of the YBCO/ITO interface in the OFF and ON states are displayed below. Solid and hollow circles respectively represent oxygen atoms and vacancies. The gradient of oxygen vacancies concentration extends over a longer length scale in the OFF state, leading to insulating, low-Tc and optimum Tc YBCO layers. Oxygen depletion exists only within a much thinner YBCO layer in the ON state. **b** Differential conductance $G \equiv dI/dV_{bias}$ plotted as a function of the applied $V_{bias}$ after a voltage pulse $V_{write}$ (~V) is first applied at T = 3.2K to set it in the ON ($V_{write} > 0$) and OFF ($V_{write} < 0$) states. The inset displays the temperature dependence of the differential conductance at zero-bias $G_0$(hollow) and at $V_{bias} = 100\ mV$ $G_{100}$(solid) in the ON (circles) and OFF (squares) states. One can see that, at zero bias ($G_0$, hollow) the conductance drops at a higher pace below a certain temperature which is close to the superconducting transition. **c** Hysteretic behaviour of the differential conductance as a function of the writing voltage $V_{write}$ for $V_{bias} = 100\ mV$ ($G_{100}$, grey) and for $V_{bias} = 0$ ($G_0$, black), which shows the maximum electrical switching amplitude $\Delta G_E$. The junction was cycled twice from positive to negative voltages, in the sense shown by the spinning arrow. **d** Relaxation of the differential conductance $\Delta G_E(t)/G_{ON}(t=0)$ as a function of time after the junction is set in the ON state($V_{write} = 6V$ at T = 3.2K), measured at different temperatures (see legend). The lines are the best fits to a stretched exponential.

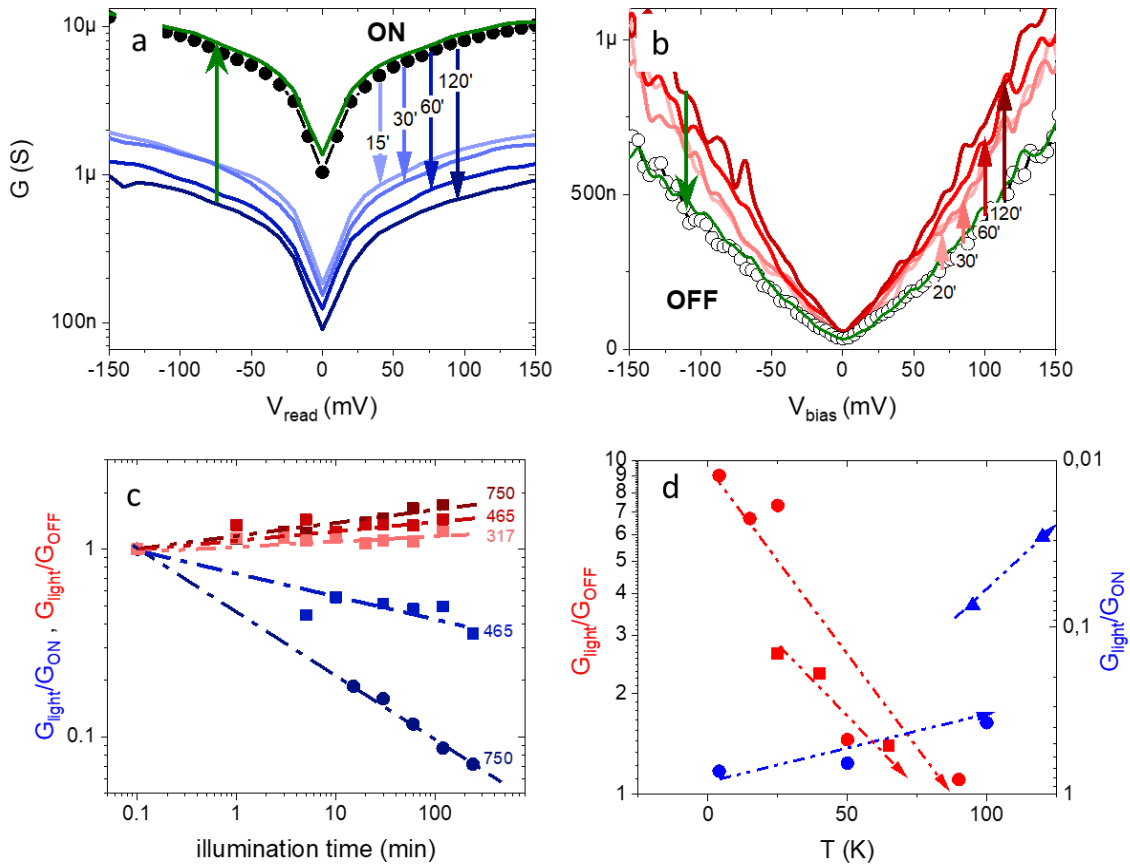

Figure 3: **Non-volatile optical switching. a-b** Differential conductance as a function of $V_{bias}$ before (black) and after illumination (colours) for increasing illumination times (see labels). In (a), the junction is set in the ON state prior to illumination, while in (b) the same sample is set in the OFF state prior to illumination. In both cases, visible light ($\lambda = 405nm$) with optical power of P = 750 $mW.cm^{-2}$ was used. $T = 100K$ for the measurements in the ON state (a) and $T = 4K$ in the OFF state (b). The temperature effects are further discussed in (d). **c** Time dependence of the light-induced relative conductance variation $G_{Light}/G_{ON}$ and $G_{Light}/G_{OFF}$ in the ON (blueish curves) and OFF (reddish curves) state respectively, for several optical powers (different colour shades) as indicated by the labels (in mW.cm$^{-2}$). Different symbol shapes (circles/squares) correspond to different junctions. In all cases, visible light ($\lambda = 405nm$) was used. $T = 100K$ for the measurements in the ON state and $T = 4K$ for those in the OFF state **d** Temperature dependence of the relative light-induced conductance variation $G_{Light}/G_{ON}$ and $G_{Light}/G_{OFF}$ for different junctions (different symbols shapes) in the ON (blue) and OFF (red) state. The illumination conditions were: visible light ($\lambda = 405nm$), with an optical power of P = 750 $mW.cm^{-2}$ in the ON state. In the OFF state, the samples were illuminated either with visible light and with an optical power of P = 150 $mW.cm^{-2}$ (red circles) or with UV light ($\lambda = 365nm$) with an optical power of P = 750 $mW.cm^{-2}$ (red squares).

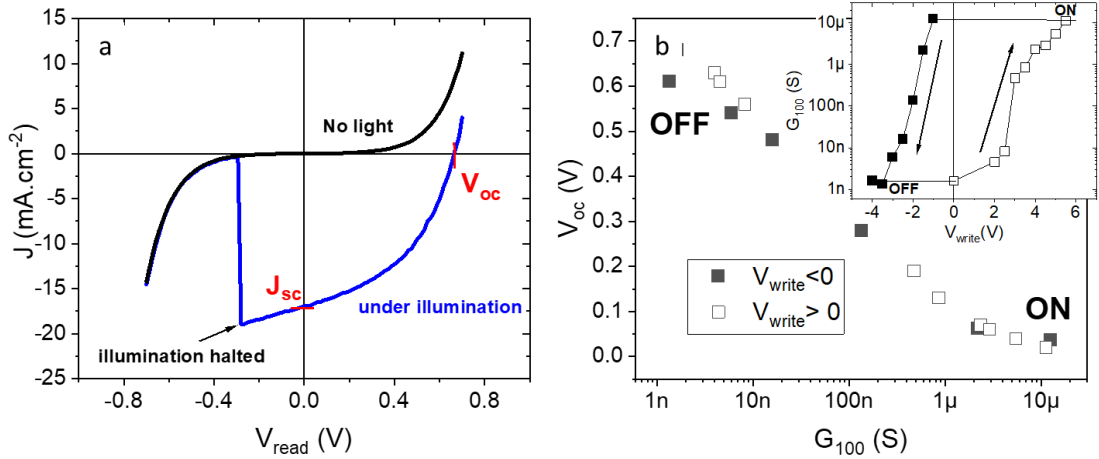

Figure 4: **Photovoltaic effect. a** Current density $J$ vs $V_{read}$ measured in dark and under illumination with visible light ($\lambda = 405nm$), with an optical power of P = 150 $mW.cm^{-2}$ and at $T = 4K$ for a junction previous set in the OFF state. The illumination is halted at the point indicated by the arrow. The open circuit voltage $V_{oc}$ and short circuit current $J_{sc}$ are highlighted. **b** Open circuit voltage $V_{oc}$ measured across the junction as a function of the remnant zero-bias conductance $G_0$ at $T = 4K$ as the junction is switched from the ON to the OFF state (solid symbols.) and the OFF to the ON state (hollow, symbols) by application of $V_{write}$. The inset shows the corresponding $G_0$ as a function of $V_{write}$.

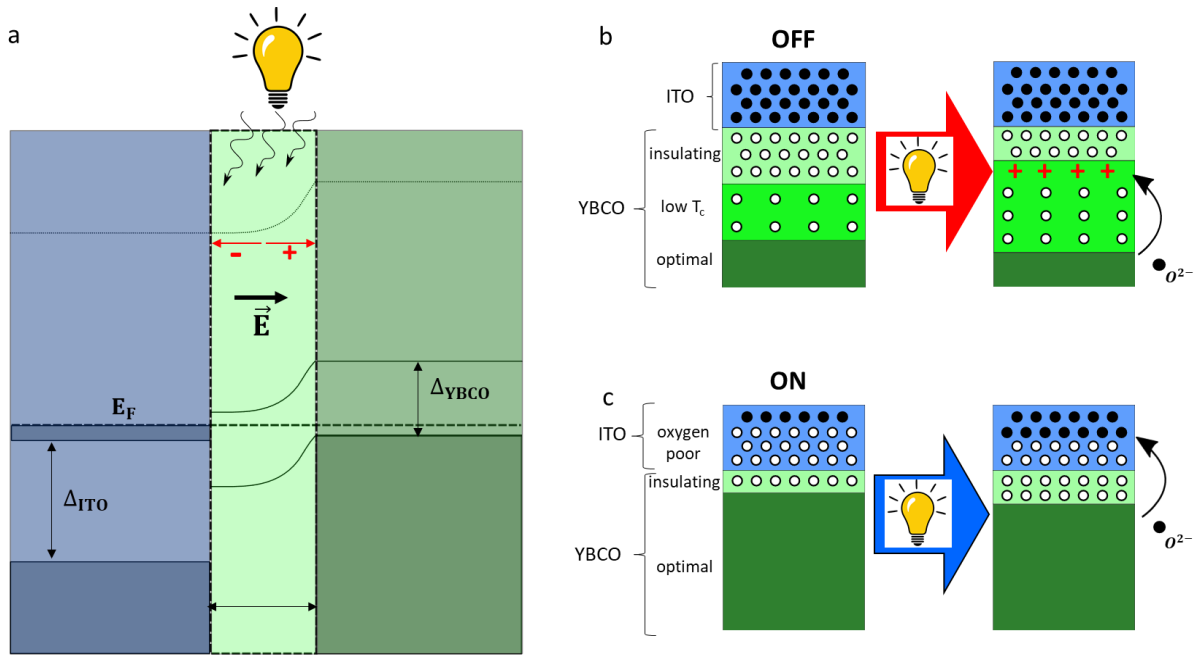

Figure 5: **Microscopic Model. a** Energy band diagram of the YBCO/ITO interface where ITO (blue) is a degenerate n type semiconductor and c-axis YBCO (green) is a p-type semiconductor. A space charge layer, characterized by a built-in electric field at the interface between the two electrodes, is indicated by the light green region. $\Delta_{ITO}$ and $\Delta_{YBCO}$ represent the electronic band gaps of ITO and YBCO respectively. **b-c** Schematic representation of the ITO/YBCO interface during optical switching. The full and hollow circles indicate oxygen atoms and vacancies respectively. In the ground state (OFF state), see **b**, the bottom electrode is highly oxygen deficient at the interface because of the high reduction potential of Cu In YBCO. Upon illumination an accumulation of holes (red crosses) occurs in the YBCO layer close to the interface which promotes the reoxidation at the interface by migration of $O^{2-}$ atom from in-depth YBCO. When the junction is set in the ON state, see **c,** oxygen has been driven away from the ITO into the YBCO yielding a conducting (excited) state for the junction. Upon illumination, the natural relaxation of the system towards its ground-state is accelerated, as light activates oxygen migration from the YBCO back to the ITO across the interface, allowing ITO re-oxidation.

## Methods

**Sample fabrication:** The 30 nm-thick c- acxis $YBa_2Cu_3O_{7-\delta}$ (YBCO) films were grown by pulsed laser depiction (PLD) on $SrTiO_3$(001) (STO) substrates, at $695°C$, in an $O_2$ atmonsphere of 0.36 mbar, using a 238nm KrF excimer laser with an energy density $\sim 1\,J.cm^{-2}$ and $\sim 5\,Hz$ repetition rate. After deposition, the films were cooled down in a pure oxygen atmosphere (800 mbar) to obtain optimal oxygen stoichiometry. Standard photolithography was used to fabricate studied junctions. First, a layer of photoresist was spin-coated unto YBCO films in which square openings (10 to200 $\mu m^2$) were patterned to define the contact area between YBCO and ITO. This first layer of photoresist was hard-baked to make it immune to further illumination and to the solvents (developers, acetone, isopropanol) used in the subsequent processing, and thus persistent. A second layer of photoresist was then spin-coated, on which larger (200 x 500 µm ) openings were defined to create electrical contact pads aligned with the small square openings in the first layer. Then the 100 nm thick ITO film was deposited by RF sputtering at room temperature, in an atmosphere mixture of argon and oxygen gas (Ar: 43 sccm/$O_2$: 3sccm), at a deposition pressure of $6x10^{-3} mbar$ and using a commercial target of $In_2O_3$ 98%/$SnO_2$2%. This was followed by a lift-off of the 2nd photoresist layer. Two wires were attached to the YBCO film (junction's bottom electrode) and one to top ITO contact pad (top electrode) using Aluminium wire and an edge bonding machine. This allows performing three-probe electrical measurements of the YBCO/ITO junctions.

**Transport measurements:** The writing voltage was applied across the junctions by ramping it from 0V up to $V_{write}$ then back to 0V. Current vs bias voltage $V_{read}$ characteristics in the remnant state were subsequently measured using a Keithley 2450 sourcemeter, with the bias across the junction $|V_{read}| \leq 200mV$ was measured using a Keithley 2182 Nanovoltmeter. The differential conductance ($V_{read}$) was obtained by numerically differentiating the measured current-voltage characteristics. In all measurements the top ITO electrode is grounded.